\documentclass{article}
\usepackage{spconf,amsmath}
\usepackage{graphicx} 
\usepackage{float}
\usepackage[dvipsnames]{xcolor}
\usepackage{multicol}
\usepackage{multirow}
\usepackage{footmisc}
\definecolor{myblue}{RGB}{115, 147, 195}
\definecolor{mygreen}{RGB}{217, 234, 211}
\usepackage{cite}
\usepackage{url}
\usepackage{boxedminipage}
\usepackage{amssymb}


\title{Longer is (Not Necessarily) Stronger: Punctuated Long-Sequence Training for Enhanced Speech Recognition and Translation}

\name{Nithin Rao Koluguri$^{1}$, Travis Bartley$^{1}$, Hainan Xu$^{1}$, Oleksii Hrinchuk$^{1}$, Jagadeesh Balam$^{1}$, Boris Ginsburg$^{1}$, Georg Kucsko$^{2}$}
\address{NVIDIA, Santa Clara, USA$^{1}$ \\ Suno Inc, Cambridge, MA, USA$^{2}$}

\email{\{nkoluguri,tbartley,hainanx,ohrinchuk,jbalam,bginsburg\}@nvidia.com, georg@suno.com}

\begin{document}
%
\maketitle
\begin{abstract}
    
This paper presents a new method for training sequence-to-sequence models for speech recognition and translation tasks. Instead of the traditional approach of training models on short segments containing only lowercase or partial punctuation and capitalization (PnC) sentences, we propose training on longer utterances that include complete sentences with proper punctuation and capitalization. We achieve this by using the FastConformer architecture which allows training 1 Billion parameter models with sequences up to 60 seconds long with full attention. However, while training with PnC enhances the overall performance, we observed that accuracy plateaus when training on sequences longer than 40 seconds across various evaluation settings. Our proposed method significantly improves punctuation and capitalization accuracy, showing a 25\% relative word error rate (WER) improvement on the Earnings-21 and Earnings-22 benchmarks. Additionally, training on longer audio segments increases the overall model accuracy across speech recognition and translation benchmarks. The model weights and training code are open-sourced though NVIDIA NeMo.
\footnotemark[1]\footnotemark[2]
\footnotetext[1]{\url{https://github.com/NVIDIA/NeMo}}
\footnotetext[2]{\url{https://huggingface.co/nvidia/parakeet-tdt_ctc-1.1b}}


\end{abstract}

\section{Introduction}

Over the past few years, we have seen great advancement in speech modeling. 'Traditional' cascade and hybrid approaches have given way to fully end-to-end neural systems. However, data preprocessing has failed to see similar advancements; it remains standard practice to casefold and strip punctuation from training corpora.  Even among recent state-of-the-art models \cite{rekesh2023fast, zhang2020pushing, radford2023robust}, training still assumes generally lowercase text (minus the rare exception for partial punctuation and capitalization). Instead of building casing and punctuation into model inference, modern systems typically leave this step to post-processing \cite{zelasko2018punctuation,pappagari2021joint}. 

Two primary factors have contributed to this:
(1) a scarcity of fully punctuated and capitalized training data, and (2) the emphasis on evaluating models on lowercase text within prominent benchmarks like LibriSpeech. While this approach has seen good performance so far, it limits the ability of such systems to leverage the rich acoustic information available in speech itself. (e.g. when a pause in speech is a strong indicator for the presence of punctuation). As such, for the next generation of speech models, there is a growing need to incorporate punctuation and casing prediction within end-to-end systems.

Further, the growing capability of modern model architectures to model long-term dependencies encourages exploration of long-duration speech input in ASR. At the time of writing, sophisticated architectures based on self-attention \cite{vaswani2017attention, gulati2020conformer,rekesh2023fast}, have shown great ability to model long context. However, the most notable gains has been in text domains \cite{an2024trainingfreelongcontextscalinglarge}. Long-form speech processing is still relatively under-explored due to constraints in computational resources as well as the intrinsic difficulty of the problem \cite{koluguri2023investigating}. Typical training samples are limited to around ~20s audio, severely hampering the ability of modern systems to learn discourse features such as conversation turns, anaphora, and tone-shifts.

In this work, we focus on exploring the FastConformer \cite{rekesh2023fast} architecture for recognition and translation speech tasks and propose several techniques to improve its performance, while training on unnormalized text, with punctuations and capitalizations. The contributions of this paper are as follows,
\begin{enumerate}
    \item We propose a new training paradigm to train models on longer utterances containing complete sentences with punctuation and capitalization, instead of short segments with partial punctuation and capitalizations.
    \item Our model employs a hybrid TDT-CTC architecture. This is the first published result using such decoders. We perform ablation studies on training models with various input sequence durations for both speech recognition and translation tasks. 
    \item By employing FastConformer, we present the first study demonstrating the feasibility of training on segments upto 60 seconds long for speech applications. 
    \item Our models brought a 25\% relative WER reduction on the Earnings-21~\cite{del2021earnings} and Earnings-22~\cite{del2022earnings} benchmarks and 15\% relative BLEU score improvement on MuST-C~\cite{di2019must} test set. 
    \item Experiments show our methods also improve model accuracy on lower-cased benchmarking sets, achieving state-of-the-art results on the HF leaderboard.
    
\end{enumerate}

The paper is organized as follows: in Section \ref{background}, we provide the background information of the main models used for our work;
in Section \ref{method}, we describe our proposed methods in detail;
in Sections \ref{datasets} and \ref{experiments}, we describe the datasets used for our work, and then the experiment results of our models on recognition and translation tasks, followed by our conclusions in Section \ref{section:conclusion}.

\section{Background}\label{background}
The models presented in this paper are based on the FastConformer \cite{rekesh2023fast} encoder trained with a Hybrid Transducer-CTC loss.

\subsection{FastConformer}
The Conformer \cite{gulati2020conformer} is a powerful architecture that comprises self-attention \cite{vaswani2017attention}, convolution, and feedforward layers. FastConformer \cite{rekesh2023fast} is a architecture that greatly reduces the Conformer computational complexity while giving superior performance with  two major changes to the original Conformer:
\begin{enumerate}
    \item A modified subsampling scheme. Instead of the commonly adopted 4x subsampling rate used for the original conformer for speech models, FastConformer's output uses an 8x subsampling rate, which cuts the output length by half. This is achieved by including 2X subsampling convolutions at the first three consecutive layers for the conformer network.
    \item The self-attention contexts of conformer blocks are redesigned to utilize a mixture of local and global attention.
\end{enumerate}

\subsection{CTC, Transducer, TDT and their Hybrid}

Transducers and CTC are two commonly used end-to-end model architectures. Both of them adopt a frame-synchronous paradigm, where the input speech signals, after being processed by the encoder, are processed frame by frame, and the model predicts a token for each of those frames. Both CTC and Transducer models adopt a blank symbol for frames that do not contribute extra information. However, CTC assumes conditional-independence between prediction and frame representations. Meanwhile, RNN-T includes a predictor network to model textual history, and a joint network to condition predictions on both frame and contextual representations.

Token-and-Duration Transducer (TDT) \cite{xu2023efficient}  is a recently proposed model that extends the traditional Transducer model by decoupling token and duration predictions. The model's duration prediction is used to skip multiple frames during decoding, thus greatly improving the inference efficiency of Transducer models. TDT models have also been shown to achieve superior accuracy than traditional Transducers, making it a good choice for our models.
\begin{figure}
    \centering
    \includegraphics[scale=0.2]{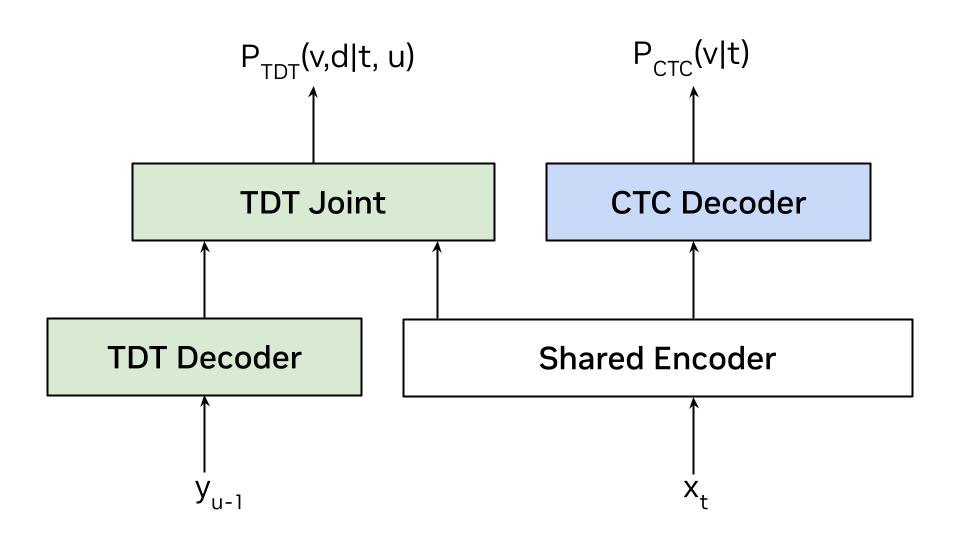}
    \caption{Hybrid-TDT-CTC Model. Variables $v$ and $d$ represent the vocabulary and durations supported by the TDT model. The final loss of the model is computed as a linear interpolation of TDT loss and CTC loss.}
    \label{hybrid_tdt_ctc_diagram}
\end{figure}
\subsubsection{Hybrid TDT-CTC models}
A hybrid Transducer-CTC model, as first utilized in \cite{noroozi2023stateful}, combines aspects of both Transducer and CTC in a single model. In this approach, the encoder output is used in two pathways: 1) combining with a Transducer prediction network and joint network, to compute a Transducer loss and 2) feeding into a CTC decoder to compute a CTC loss. The two losses are combined with weighted summation to give the final loss for training. Thanks to this dual loss objective, hybrid models effectively train two models at once. Along with their faster convergence rate (in comparison to constituent models), hybrid models are also highly compute efficient. Needing to conduct several ablations over the course of this work, we chose hybrid models as our primary architecture.

Further, we propose a new hybrid-TDT-CTC architecture, shown in Figure \ref{hybrid_tdt_ctc_diagram}. The model replaces the Transducer components \cite{noroozi2023stateful} with TDT.  TDT  offers two key advantages:  1) TDT models employ an architecture similar to RNN-T yet have demonstrated superior performance on various speech tasks, and 2) TDTs achieve significantly greater inference speed than RNN-T models, making them well-suited for deployment in both cloud and on-device computing environments.  For computing the Hybrid-TDT-CTC model loss, we have,
$$
\mathcal{L}_\text{final} = \mathcal{L}_\text{TDT} + \lambda \mathcal{L}_\text{CTC}
$$
By this design, the CTC decoder and TDT model's decoder and joint network are independently trained, while the encoder receives gradients from both objectives. This approach allows the CTC decoder to converge faster\cite{noroozi2023stateful} while resulting in better time alignments for both the shared decoders\cite{sudo20244d}.

\vspace{-4pt}
\section{Method} \label{method}

In this section we overview  sentence-level and long audio training  with punctuation and capitalization.  

\subsection{Sentence-Level Training with Punctuation and Capitalization (PnC)}

Previous works, such as FastConformer~\cite{rekesh2023fast}, Conformer~\cite{zhang2020pushing, meister2023librispeech}, and Whisper~\cite{radford2023robust}, have reported training on segments containing partial punctuations and capitalizations or lower-case ground truth text samples. Our proposed method advocates for concatenating partial segments to form complete sentences originating from the same utterance. "Complete sentence" in this context refers to a sentence beginning with a capital letter and concluding with a punctuation mark (period, exclamation point, or question mark).

We illustrate the application of our PnC method to segments from LibriSpeech-PC \cite{meister2023librispeech} set, showcasing the combined sentences in figure \ref{fig:highlighted_sentences}.

\begin{figure}[t]
\centering
\begin{boxedminipage}{\linewidth}
\textbf{Previous:}

\texttt{102-129232-0076}: \color{OliveGreen} {What appears once in the atmosphere may appear often, and it was undoubtedly the archetype of that familiar ornament. I have seen in the sky a chain of summer lightning,} \\
\color{Black}
\texttt{102-129232-0077}: \textcolor{myblue} {which at once showed to me that the Greeks drew from nature when they painted the thunderbolt in the hand of Jove.} \\
\color{Black}
\textbf{After (Combined Sentence):}

\texttt{102-129232-0076\_0077}: \color{OliveGreen} What appears once in the atmosphere may appear often, and it was undoubtedly the archetype of that familiar ornament. I have seen in the sky a chain of summer lightning, \textcolor{myblue}{which at once showed to me that the Greeks drew from nature when they painted the thunderbolt in the hand of Jove.}
\end{boxedminipage}
\caption{Method of concatenating partial punctuations and capitalizations segments from LibriSpeech-PC \cite{meister2023librispeech} set to form a complete sentence level segments.}
\label{fig:highlighted_sentences}
\end{figure}
\color{Black}
\subsection{Training with Longer Context}

As exemplified by the sentence formation process described above, the proposed method requires training on longer audio segments.  To handle these extended utterances, the FastConformer architecture is employed due to its efficient conformer modules and downsampling module that reduces the input sequence size by a factor of 8, enabling efficient processing of longer sequences. These characteristics enable the model to be trained on 60-second audio segments on NVIDIA A100 80GB GPUs, compared to the 30 seconds or less limit observed in previous works \cite{radford2023robust,  fox2023updated}, leveraging the benefits of longer context for improved performance as demonstrated in the experiments section.

\section{Datasets} \label{datasets}
This section details the datasets employed for training and evaluating the proposed methods.

\subsection{Speech Recognition}
\subsubsection{Training}
\label{sec:training-data}
Due to the limited availability of large-scale PnC data, we leverage a combination of internally curated data and publicly available datasets consisting of samples from LibriSpeech-PC \cite{meister2023librispeech}, Fisher \cite{cieri2004fisher}, MCV-11 \cite{ardila2019common}, MLS \cite{pratap2020mls}, NSC Part1 \cite{tanspontaneous}, SPGI \cite{o2021spgispeech}, VCTK \cite{veaux2017vctk}, VoxPopuli \cite{wang2021voxpopuli}. Dataset statistics for combined data are provided in Table \ref{table:asr-training-set}. For experiments on segments consisting of partial punctuations and capitalizations, we employ full data from public datasets, and for experiments on complete sentence training that starts with an upper case letter and ends with punctuation (.!?), as illustrated in Figure \ref{fig:highlighted_sentences}, we use only those segments from the datasets.
\begin{table}[t]
    \caption{Statistics of combined internal and public  datasets with partial and improved punctuation and capitalizations sets. }
    \vspace{0.15cm}
    \label{table:asr-training-set}
    \centering
    \resizebox{0.47\textwidth}{!}{%
        \begin{tabular}{c|c|c|c}
            \hline 
            \begin{tabular}{c} 
                Complete \\
                Sentences
            \end{tabular} & 
         \begin{tabular}{c} 
            Duration \\
            Window (sec)
            \end{tabular} & 
            \begin{tabular}{c}
                 Duration\\
                 (hrs)
            \end{tabular} & 
            \begin{tabular}{c} 
                Avg Segment \\
                Duration (sec)
            \end{tabular}\\
            \hline
            \hline   $\times$ & $0-20$ & 18,784.31& 9.13\\
            \hline   $\checkmark$ & $0-20$ &17,983.04& 11.03\\
            \hline   $\checkmark$ & $20-40$ &7,123.28& 33.48\\
            \hline   $\checkmark$ & $40-60$ & 5,557.22& 51.49\\\hline
        \end{tabular}%
        }

\end{table}
\subsubsection{Evaluation}
Prior research on PnC for ASR, such as LibriSpeech-PC, utilized evaluation sets derived from the LibriSpeech dev and test sets. However, these segments deviate from typical conversational speech as they are truncated and lack complete sentence structure as shown in Figure \ref{fig:highlighted_sentences}. Therefore, we employ the Earnings-21 and Earnings-22 datasets for model evaluation.

Earnings-21 and Earnings-22\footnote{\url{https://github.com/revdotcom/speech-datasets}} are collections of earnings calls spanning various financial sectors. Earnings-21 comprises 39 hours of audio, while Earnings-22 offers 119 hours. Both datasets are commonly used to benchmark ASR systems for long-form audio transcription. Previous studies \cite{koluguri2023investigating} evaluated these datasets with normalized text and lowercase letters. In contrast, we restore punctuations and capitalizations from reference files to assess model performance under more realistic conditions.

Given the hour-long nature of utterances in Earnings-21 and Earnings-22, and the 22-minute inference limit of the FastConformer TDT XXL model\cite{rekesh2023fast} with full self-attention module, we leverage the NeMo NFA\footnotemark[1] tool with the FastConformer CTC XXL\cite{rekesh2023fast} model to generate word-level timestamps. These timestamps, combined with punctuation, are then used to segment the audio files into 20-minute approximate chunks. We then use NeMo text normalization\footnotemark[1] toolkit to convert numerical numbers to their corresponding text form. The resulting segmented data is used for evaluating the models and is summarized in the table \ref{tab:earnings-eval-set}.

\footnotetext[2]{\url{https://github.com/NVIDIA/NeMo}}

\begin{table}[t]
    \caption{Earnings-21/22 evaluation dataset statistics. }
    \vspace{0.15cm}
    \label{tab:earnings-eval-set}
    \centering
    \resizebox{0.47\textwidth}{!}{%
        \begin{tabular}{c|c|c|c}
            \hline \multicolumn{1}{c}{ Dataset } & \multicolumn{1}{|c|}{ \# Segments } 
            & \begin{tabular}{c} 
            Mean segment \\
            Duration  \\
            (mins)
            \end{tabular} & \begin{tabular}{c} 
            Std of \\
             durations \\
            (mins)
            \end{tabular} \\
            \hline
            \hline Earnings-21 & 138 & 17.06 & 5.46  \\
            \hline Earnings-22 & 430 & 16.72 & 6.08 \\
            \hline
        \end{tabular}%
        }
\end{table}


\subsection{Speech Translation}
Speech translation experiments leverage the German subset of the MuST-C dataset \cite{di2019must}. This multilingual corpus features audio-aligned translations of TED Talks, offering diverse training samples ranging from one second to over a minute in duration. With multiple samples originating from the same talk, MuST-C allows for the concatenation of sequential samples to create longer audio sequences. This characteristic allows use of our previous method for the task of speech translation.

We establish a baseline subset from the ``original'' dataset, containing only audio samples with durations ranging from 0 to 20 seconds. To evaluate the impact of ``extended" context lengths, we produce additional partitions of the training set by greedily concatenating sequential audio samples up to a maximum target duration of 20, 40, and 60 seconds. Table \ref{table:ast-training-set} summarizes the total average durations for each partition. 

For evaluation, we only create greedy partitions of the original development and test sets up to 60s in length. As both subsets only contain samples of maximum length $\sim$60s, these partitions effectively contain all audio from the original subsets.

\begin{table}[t]
    \caption{Statistics of the MuST-C German translation training set: the number of hours in each duration window bucket.}
    \vspace{0.15cm}
    \label{table:ast-training-set}
    \centering
    \resizebox{0.47\textwidth}{!}{%
    \begin{tabular}{c|c|c|c}
        \hline
        \begin{tabular}{c}
         Duration \\
         Context 
        \end{tabular} & 
        \begin{tabular}{c} 
        Duration \\
        Window (sec)
        \end{tabular} & \begin{tabular}{c} 
        Duration \\
        (hrs)
        \end{tabular} & \begin{tabular}{c} 
        Avg. Segment \\
        Duration (sec)
        \end{tabular} \\
        \hline
        \hline
         Original  & $0-20$ & 402.86 & 5.58 \\
        \hline & $0-20$ & 402.86 & 5.79 \\
        \cline { 2 - 4 } Extended & $0-40$ & 430.13 & 6.11 \\
        \cline { 2 - 4 } & $0-60$ & 432.93 & 6.14 \\
        \hline
    \end{tabular}%
    }
\end{table}

\section{Experiments \& Results} \label{experiments}

\begin{table*}[t]
    \centering
    \caption{The performance of FastConformer XXL models on HF-ASR leaderboard test sets~\cite{ami,del2021earnings,chen2021gigaspeech,panayotov2015librispeech,o2021spgispeech,hernandez2018ted,wang2021voxpopuli,ardila2019common}. The table illustrates the performance improvement (WER reduction) of models with complete sentence PnC data and longer context duration samples on non-PnC datasets. Evaluations were done with TDT decoder. Greedy WER (\%) after whisper normalization~\cite{radford2023robust}. }
    \vspace{0.15cm}
    \resizebox{\textwidth}{!}{%
    \label{tab:hf-leaderboard}
    \begin{tabular}{c|c|c|c|c|c|c|c|c|c|c|c }
        \hline \begin{tabular}{c} 
        Complete \\
        Sentences
        \end{tabular} & Duration window & AMI & Earnings-22 & Giga Speech & LS clean & LS other & SPGI & TEDLIUM-V3 & VoxPopuli & MCV 8.0 & Avg WER \\
        \hline
        \hline $\mathbf{x}$ & $0-20$ & 16.21 & 11.78 & 9.99 & 1.62 & 3.16 & 1.92 & 3.42 & 5.98 & 8.68 & 6.95* \\
        \hline$\checkmark$ & $0-20$ & 15.62 & 11.35 & 10.05 & 1.74 & 3.1 & 2.05 & 3.9 & 5.86 & 7.71 &6.82 \\
        \hline$\checkmark$ & $0-40$ & 14.94 & 11.49 & 9.99 & 1.69 & 3.09 & 2.05 & 3.92 & 5.93 & 7.49 &\textbf{6.73} \\
        \hline$\checkmark$ & $0-60$ & 14.92 & 11.58 & 10.17 & 1.79 & 3.31 & 2.14 & 3.83 & 6.01 & 7.57 &6.81 \\
        \hline
    \end{tabular}%
    }
    \vspace{-0.35cm}
\begin{flushright}
\small *current top of the leaderboard parakeet\_tdt\_1.1b model.
\end{flushright}
\end{table*}

\subsection{Speech Recognition}
We investigate the impact of training with improved punctuation and capitalization training with complete sentence-level data and longer duration (context) length on ASR performance.

\subsubsection{Sentence-Level Training}
We fine-tuned the pre-trained FastConformer XXL model (1.1B parameters \cite{rekesh2023fast}) with hybrid TDT-CTC decoder on the combined dataset.
The CTC-hybrid architecture is chosen because it has been shown to have better convergence rates, and better time alignments since the CTC loss can help the Transducer learn alignment much faster during training \cite{yan2022ctc}; and TDT since it achieves similar accuracy but significantly faster inference than RNN-Transducers \cite{xu2023efficient}.
To evaluate the effectiveness of sentence-level training, we compared two subsets. The first subset contained short segments with partial punctuations and capitalizations, while the second subset comprised complete sentences. For both subsets, we limited the samples to 0-20 seconds during training. We selected only those utterances common to both subsets. While this approach may result in a different number of utterances, the total duration for both subsets remain equal. We finetuned both models for 25,000 steps with initial warmup of 5000 steps and max learning rate of 3e-4 using AdamW optimizer, Inverse Square Annealing scheduler, and hybrid TDT-CTC loss with CTC weight ($\lambda$) of 0.3 before adding to TDT-loss. All experiments are carried on 8 nodes with each node consisting of 8 GPUs.  We evaluated performance across four settings: presence/absence of punctuations and capitalizations on the Earnings-21 and Earnings-22 datasets as shown in Table \ref{tab:earnings-21-results}.

Results show a relative improvement of $\sim$20\% on Earnings-21 set and $\sim$22\% on Earnings-22 solely from training with complete sentences for punctuation and capitalization evaluations. We also observe improvement on non-punctuated/non-capitalized evaluations. This suggests that training with complete sentences helps attention mechanisms learn sentence and language structure better than training with partial segments.

\subsubsection{Context Length Extension} 
\label{sec:context-length-extenstion}
 We further investigated the impact of using longer sequences during training on the model's learning capability. To do this, we included additional samples from the training datasets to prepare two more bins of 20-40 sec and 40-60 sec duration window segments. We then fine-tuned two models with these added duration bins from 0-40 sec and 0-60 sec, and evaluated their performance using the same settings as before. As with earlier models, we fine-tuned these models for 25,000 steps, using the same hyper-parameters as the previous experiment. In these runs, we were able to fit samples of 60 sec segment buckets with a batch size of 2, and 40 sec segment buckets with a batch size of 8 on NVIDIA A100 80GB GPU with a one billion parameter model.

Consistent with prior work\cite{fox2023updated}, increasing the training segment length up to 40 sec improved performance by an additional 5\% relative reduction in WER. Training with complete sentences with longer context resulted in an overall relative improvement of approximately 25\% over partial segments with punctuations and capitalizations. However, when we trained with additional segments from 40-60 sec duration bins, we found the model accuracy plateaued with minor decrease in WER with PnC, while showing a slight increase in WER for remaining evaluation settings.

To evaluate the impact of context length during training with our proposed data preparation method, and to determine whether it's a model limitation or an effect of data preparation, we trained an additional model with complete (extended) sentences but without PnC using the same 0-60 sec extended duration windows (see table: \ref{tab:earnings-21-results}). This model was trained identically to the previous one, with the same number of steps and optimizer. As shown in table \ref{tab:longer-context}, the model trained with PnC outperformed the model trained without PnC when evaluated on the Earnings 21 and Earnings 22 lower case sets inferring that PnC model training with complete sentences learns better with added semantic information. However, the slight performance difference suggests that training on longer segments with PnC on complete sentences can be beneficial, but the effectiveness is capped by the model's ability to learn from extended segments. We hypothesize this may be due to the attention layers ability to attend to longer sequence lengths. We leave this for a future work on improving the architecture for effective training on very long sequences in speech domain. 


Furthermore, we evaluated our models on the HuggingFace ASR leaderboard \footnote[3]{\url{https://huggingface.co/spaces/hf-audio/open_asr_leaderboard}} benchmark sets. As shown in Table \ref{tab:hf-leaderboard}, the proposed data preparation techniques achieved SOTA results on lowercase evaluation sets, showcasing the model's performance on all benchmark sets. These findings demonstrate the effectiveness of sentence-level training and longer context segments in improving ASR accuracy, even generalizing to unseen scenarios like non-punctuated/non-capitalized evaluations. As observed before, we can see that the best WER is achieved with model trained on durations up to 40 seconds.

\begin{table}[t]
    \caption{Comparison of WER \% ($\downarrow$) on Earnings-21/22 for models trained with and without punctuation and capitalization for various sequence durations, evaluated with the TDT decoder.}
    \vspace{0.15cm}
    \label{tab:earnings-21-results}
    \centering
    \resizebox{0.47\textwidth}{!}{%
    \begin{tabular}{@{} c @{} c @{} c @{} c c c}
        \hline \begin{tabular}{c} 
        Complete\\
        Sentences
        \end{tabular} & \begin{tabular}{c} 
        Duration\\
        Window (sec)
        \end{tabular} & PnC & Only Cap & Only Pun & No PnC \\
        \hline 
        \hline
        & &\textbf{\textit{Earnings-21}}& & & \\
        \hline  $\times$ & $0-20$ & 24.86 & 22.29 & 15.49 & 12.44 \\
        \hline  $\checkmark$ & $0-20$ & 19.96 & 17.97 & 14.28 & 12.12 \\
        \hline  $\checkmark$ & $0-40$ & 18.98 & \textbf{17.01} & \textbf{13.64} & \textbf{11.61} \\
        \hline $\checkmark$ & $0-60$ & \textbf{18.86} & 17.09 & 13.65 & 11.63 \\
        
        \hline
         & &\textbf{\textit{Earnings-22}}& & & \\
        \hline  $\times$ & $0-20$ &30.19& 27.00 & 19.41& 15.6\\
        \hline  $\checkmark$ & $0-20$ & 23.58& 21.43& 17.06& 14.72\\
        \hline  $\checkmark$ & $0-40$ & 22.62& \textbf{20.35}& \textbf{16.35} & \textbf{14.13}\\
        \hline  $\checkmark$ & $0-60$ & \textbf{22.4}& 20.54 & 16.38 & 14.15\\
        \hline
  
    \end{tabular}%
    }
\end{table}

\begin{table}[h]
    \caption{Comparison of WER \% ($\downarrow$) for models trained with and without PnC using complete sentences on Earnings-21/22 lowercase text, evaluated with the TDT decoder.}
    \vspace{0.15cm}
    \label{tab:longer-context}
    \centering
    \resizebox{0.47\textwidth}{!}{%
        \begin{tabular}{|c|c|c|c|c|}
        \hline \begin{tabular}{c} 
        Duration \\
        Window
        \end{tabular} & \begin{tabular}{c} 
        Complete \\
        Sentences
        \end{tabular} & \begin{tabular}{c} 
        PnC during \\
        training
        \end{tabular} & Earnings 21 & Earnings 22 \\
        \hline $0-60$ & $\checkmark$ & ${\checkmark}$ & \textbf{11.63} & \textbf{14.15} \\
        \hline $0-60$ & $\checkmark$ & $\times$ & 11.71 & 14.69 \\
        \hline
        \end{tabular}%
    }
\end{table}

\subsubsection{Comparison with Cascaded Models}

To evaluate the effectiveness of training with punctuation and capitalization on complete sentences, we also assessed models trained with 0-60 seconds of complete sentences but without punctuation and capitalization, as described in subsection \ref{sec:context-length-extenstion}, and compared them with the proposed PnC training model. To obtain punctuation and capitalization from the lower-case predicted baseline model text, we used an open-source text-to-text PnC model based on a BERT encoder\footnote[4]{\url{https://catalog.ngc.nvidia.com/orgs/nvidia/teams/nemo/models/punctuation_en_bert}}\cite{devlin2018bert} trained on the Librispeech\cite{panayotov2015librispeech} and Fisher text corpora\cite{cieri2004fisher}. As shown in Table \ref{tab:cascaded-models}, the model trained with the proposed method performs better than the cascaded baselines on both the earnings 21 and earnings 22 evaluation sets. This result also highlights the necessity of end-to-end models for transcribing with punctuation and capitalization (PnC). Using an acoustic encoder allows the model to capture both the acoustic and semantic meanings of text, enabling the generation of correct punctuation and capitalization.

\begin{table}[t]
    \caption{Comparison of WER \% ($\downarrow$) on Earnings-21/22. The baseline model was trained with lower-case text, and punctuation and capitalization were later applied using a cascaded external text-to-text PnC BERT model. Evaluated with TDT decoder.}
    \vspace{0.15cm}
    \label{tab:cascaded-models}
    \centering
    \resizebox{0.47\textwidth}{!}{%
    \begin{tabular}{@{} c |@{} c @{} c @{} c c c}
        \hline \begin{tabular}{c} 
        Model 
        \end{tabular} & \begin{tabular}{c} 
        External Text \\
        PnC
        \end{tabular} & PnC & Only Cap & Only Pun \\
        \hline 
        \hline
        & &\textbf{\textit{Earnings-21}}& & & \\
        
        \hline 
        \multirow{2}{*}{FastConformer-XXL w/o PnC}   & No & 29.43 & 23.34 &  19.36\\
        \cline{2-5}  & Yes & 23.44 & 19.60 &  16.20 \\
        \hline FastConformer-XXL w/ PnC & No & \textbf{18.86} & \textbf{17.09} & \textbf{13.65}  \\
        
        \hline
         & &\textbf{\textit{Earnings-22}}& & & \\
        
        \hline
        \multirow{2}{*}{FastConformer-XXL w/o PnC}   & No & 32.41 & 26.48 & 21.65 \\
        \cline{2-5} & Yes & 26.71 & 22.76 & 18.81 \\
        \hline FastConformer-XXL w/ PnC & No & \textbf{22.4}& \textbf{20.54} & \textbf{16.38}  \\
        \hline
  
    \end{tabular}%
    }
\end{table}

\vspace{-3pt}
\subsection{Speech Translation}
We extend the prior investigation by analyzing the effect of increased context duration and also comparison of CTC and TDT decoders with hybrid models on speech translation performance. We employed the FastConformer Large model (120M parameters) for speech translation model training. The encoder weights were initialized from a FastConformer RNN-Transducer ASR model, following \cite{rekesh2023fast}. Models were fine-tuned on the MuST-C training set for 200k steps using Adam optimizer with a learning rate of $1e-3$ and inverse-square-root annealing (warm-up 2500 steps) on 16 NVIDIA A100 80GB GPUs. We use a vocabulary size of 16000 trained using unigram tokenization \cite{kudo2018subword} over both the MuST-C German corpus and internal datasets.

To address varying speech lengths, data was segmented into 10-second intervals with corresponding batch size adjustments: 16 for [0,10), 8 for (10,20], (20,30], and (30,40], and 4 for (40,50] and (50,60]. Additionally, gradients were accumulated over two batches per training step. We evaluated performance on MuST-C \cite{di2019must} development and test sets using both CTC and TDT decoding modules.

These models are trained with PnC enabled for all experiments and similar to improvements in speech recognition WER, incorporating a longer context window led to enhanced translation quality in speech translation. However, as shown in Table \ref{tab:must-c-combined}, a longer context window upto 40 seconds yielded superior performance compared to the baseline dataset across all subsets. While this improvement could be partially attributed to the increased dataset size for the 40-second and 60-second context window subsets, it is important to note that the 20-second context window subset utilized the same amount of duration hours as the baseline. Therefore, the observed performance gains can be directly ascribed to the extended context window. However, these gains diminish after a context window of 40 seconds, which is similar to our observation in speech recognition experiments as well. 

\begin{table}[t]
    \caption{Comparison of BLEU ($\uparrow$) scores between models trained with longer context / durations and models trained with the original MUST-C sample durations.}
    \vspace{0.15cm}
    \label{tab:must-c-combined}
    \centering
    \resizebox{0.45\textwidth}{!}{%
    \begin{tabular}{c|c|c|c|c}
        \hline 
        \begin{tabular}{c} 
        Context \\
        Length
        \end{tabular} &
        \begin{tabular}{c} 
        Duration \\
        Window (sec)
        \end{tabular} & 
         Decoding & Validation & Test \\
         \hline
        \hline  \multirow{2}{*}{ original } & \multirow{2}{*}{$0-20$} & CTC &  20.17 & 18.92 \\
         & & TDT & 22.1 & 21.53 \\
        \hline  \multirow{6}{*}{ extended } &\multirow{2}{*}{$0-20$} & CTC & 21.23 & 20.03 \\
         & & TDT & 23.09 & \textbf{22.42} \\
         \cline{2-5}  & \multirow{2}{*}{$0-40$} & CTC & \textbf{21.26} & \textbf{20.45} \\
         & & TDT & \textbf{23.12} & 22.23 \\
         \cline{2-5}  & \multirow{2}{*}{$0-60$} & CTC & 20.88 & 19.90 \\
         & & TDT & 22.5 & 21.92 \\
        \hline
    \end{tabular}%
    }
\end{table}
\section{Conclusions}\label{section:conclusion}

In this paper, we introduce a novel approach to training punctuation and capitalization models using complete sentences. By utilizing the FastConformer architecture and extending training segment durations from 20 seconds to 60 seconds, we present the first study demonstrating the feasibility of training on segments upto 60 seconds long. In this work, to ensure faster convergence we employed a hybrid TDT-CTC loss. Our method achieved a significant relative improvement of 25\% on the Earnings-21/22 evaluation datasets and additionally, it enhanced performance on short-duration, lower-case benchmarks. This method, when applied to speech translation, yielded a 15\% relative BLEU score improvement on the MuST-C test set. While our findings highlight the general benefits of longer context training, we observed diminishing returns for segments exceeding 40 seconds when evaluated on shorter duration benchmarks in both speech recognition and translation tasks. In future work, we plan to explore effective training on very long sequences with an improved architecture. We believe this research paves the way for training on longer audio sequence lengths in other large seq2seq models, including those used for speaker diarization and music applications.

\bibliographystyle{IEEEbib}
\bibliography{mybib}

\end{document}